\documentclass{article}

\usepackage{arxiv}

\usepackage[utf8]{inputenc} 
\usepackage[T1]{fontenc}    
\usepackage{hyperref}       
\usepackage{url}            
\usepackage{booktabs}       
\usepackage{amsfonts}       
\usepackage{nicefrac}       
\usepackage{microtype}      
\usepackage{lipsum}
\usepackage{graphicx}

\usepackage[T1]{fontenc}

\usepackage[table,dvipsnames,xcdraw]{xcolor}  
\usepackage{tabularx}
\usepackage{booktabs}
\usepackage{enumitem}

\usepackage{pdflscape}  
\usepackage{afterpage} 
\usepackage{float}  
\usepackage{placeins} 

\usepackage{graphicx}
\usepackage{pgfgantt}
\usepackage{xcolor}
\usepackage[normalem]{ulem} 
\usepackage{color,soul} 

\usepackage{multicol}
\usepackage{graphicx}

\usepackage{subfigure}


\title{GrADyS: Exploring movement awareness for efficient routing in Ground-and-Air Dynamic Sensor Networks}

\author{
 Bruno José Olivieri de Souza \\
  PUC-Rio\\
  \texttt{bolivieri@inf.puc-rio.br} \\
   \And
 Marcelo Paulon Jucá Vasconcelos \\
  PUC-Rio\\
  \texttt{mvasconcelos@inf.puc-rio.br} \\
  \And
 Markus Endler \\
  PUC-Rio - Pontifícia Universidade Católica Do Rio De Janeiro\\
  Rio de Janeiro - Brasil\\
  \texttt{endler@inf.puc-rio.br} \\
}

\begin{document}
\maketitle
\begin{abstract}
Several situations exist where a geographic region of some size needs to be scanned or monitored through many sensors. Still, it is either absolutely impossible or prohibitively expensive to deploy and maintain wireless communication infrastructure for the distributed sensors. Either because the region is hidden behind walls, not easily accessible, hard to get through, or infected with some lethal bacteria or virus transmitter.  
In this case, the best is to scatter (disposable) sensors in the region and let them transmit the collected sensor data by wireless means to an overflying UAV/drone. Which then physically hauls the collected data from the monitored area to a central base station that functions as a gateway to the Internet.  The project GrADyS aims to research two sets of problems regarding such data collection. The former aims to coordinate several autonomous UAVs in a distributed manner to collect the generated data while relying only on ad-hoc communication. The latter aims to develop routing protocols to mesh networks Bluetooth Mesh's Low Power Nodes. Both research lines already present preliminary results that are presented in this paper.
\end{abstract}


\section{Introduction}

This research is composed of two - mutually supportive - parallel projects. The first project, named GrADyS-A (GrADyS AIR), is focused on the task of collect sensor data from a WSAN installed on the ground by a fleet of UAVs aiming to maximize the throughput of the soil data collection. The assumption is that these UAVs fly independently over the monitored area and use only short-range radios for Peer-to-Peer wireless communication, both for accessing the sensors on the ground and swapping data with nearby UAVs.
Thus, each UAV only exchanges data when they eventually meet (i.e., get close to) another UAV or when they approach Ground Station, connected to the Internet. At each meeting between UAVs, they will also change their flight plans (itinerary) for further data collection. 

The GrADyS-A of this work builds on previous research documented in \cite{olivieriIROS17}\cite{olivieriMSWIM17}. In this solution, the bottleneck is the number of UAVs, since a single UAV may miss sensor data from the sensors spread in the region while carrying data back and forth. Therefore, in situations where one needs continuous environmental monitoring of a region with the smallest possible loss of data, it is mandatory to use a fleet of roaming UAVs that coordinate their flight trajectories and their data swap operations to minimize the intervals of time where sensors on the ground are left unattended.

In terms of basic research, GrADyS-A evolves and refines the proposed distributed algorithm for UAV fleet coordination and communication so to optimize UAV relative positioning and maximize data collection throughput for different situations of sensor distributions and flight distance to/from the base station.  We also use new performance analysis methods and metrics to assess and compare different coordination algorithms—many of these assessments through simulations.

An additional goal of the GrADyS-A is to conduct applied research using the proposed coordination algorithms and their performance characteristics.  In this regard, a small fleet of UAVs (i.e., quadcopters) will be used to carry out tests in the field similar to the simulated use cases and scenarios. These real-world tests will provide evaluations of results obtained and also identify new real-world aspects that could affect data collection by the fleet of UAVs.

The UAV fleet coordination goal is to maximize the throughput and keep uninterrupted data collection from WSAN ground nodes capable of transmitting data upwards using a Bluetooth Low Energy. The GrADyS-A of our work's main originality lies in the decentralized flight coordination approach using a lightweight P2P protocol among the UAVs that we believe will be the cornerstone of future self-controlled, flexible and efficient multiple-UAV area scanning procedures and interactions with dynamic ground-based networks (WSANs). Moreover, there is very little research work on mobile ad hoc network protocols (including {\em Flying Ad Hoc Nets - FANETs}) that combine and compare performance analysis both through simulation and {\em tests {\em in the wild}}.
Our approach of decentralized UAV coordination (for movements and communications) will be explained in further detail in section \ref{L:part1}.

In the second member project, named GrADyS-G, we use the recent Bluetooth Mesh (BT Mesh) protocol standard to investigate energy-efficient and dynamic means of data routes from each WSAN node to mobile data collectors (e.g., mobile nodes or UAVs) with minimum global energy expenditure utilizing Bluetooth Mesh's concepts of Friend Node (FN) and Low Power Nodes (LPN).\cite{Bluetooth:2019}.
In terms of basic research, we are designing and implementing (in the OMNeT++ INET simulator framework\cite{INET-www}) a flexible and dynamic information routing protocol for the BT Mesh, to haul "fresh" data collected from a full set of mesh nodes to moving collector/sink nodes with short-lived connectivity.

Furthermore, this flexible routing uses opportunistic in-network buffering and data aggregation functions and will use AI for the WSAN Mesh network to "learn" and anticipate the moving sink/collector nodes' future connection. Much of the experimentation and analysis of the routing protocols are simulation-based, as it is easier to test different mesh network sizes,  sensor data generation frequency, different densities of Friend Nodes, and different movement behaviors of the collector/sink nodes.  So far, we have already finished a basic version of the INET-based routing protocol and are beginning to measure the protocols' efficiency for some Mesh network and (mobile) collector node settings. 

In addition to the above, GrADyS-G (GrADyS Ground) involves applied research, where we will evaluate the developed routing protocol in a real-world situation, by instrumenting several trees of the University campus with  ESP32 SoC (as BT mesh nodes) with humidity and SAP sensors and test the efficiency of the data collection to passing pedestrians and some UAV overflying the group of trees. 
Hence, GrADyS-G leads to the design and evaluation of a novel and practical Mesh information routing protocol "for a ground Mesh network" - entirely compatible with the Bluetooth Mesh routing\cite{BT-Mesh-www}\cite{Junjie:TSN:2019} standard.  This proposed protocol will employ activeness and mobility-awareness for routing data from sensors towards mobile collector nodes and aiming the most efficient transfer of data from the ground to whichever mobile/flying node happens to be visiting the Mesh network.
It is thus a perfect complement to the research in GrADyS-A.

The main innovation in the GrADyS-G, is that it develops and test a routing approach that can handle data delivery not only to one but to several mobile data collectors at the same time, maximizing the data throughput while at the same time coping with the periodic and short-time "awakening" of Bluetooth Mesh's Low Power Nodes. Moreover, as we create a protocol that extends part of the routing, we aim to make it applicable to the other IoT devices, SmartPhones, and other devices that could interact with the network.

Considered together, the GrADyS-A and GrADyS-G projects focus on the information layer for dynamic routing in (ground-to-air) networks that operate in tandem and where the main common characteristics are node mobility and opportunistic, short-lived wireless interactions. In both cases, this project designs, simulate, implement on real hardware/UAVs, and evaluate the proposed protocols in real-world scenarios.

Therefore, with this project, we believe to be paving the road towards the development of modern mobile environmental, wildlife, and surveillance systems and pioneering the area of flexible and efficient interaction between mobile ground and flying networks.

\section{GrADyS-A}
\label{L:part1}

The use of unmanned aerial vehicles (UAV), otherwise known as \textit{drones}, for several purposes, has significantly increased in the last decade \cite{Reina2018a}. UAVs offer agile and cost-effective solutions for many demanding military and civilian applications \cite{Al-Hourani2018}, and have drawn significant research interest in recent years due to their wide range of applications, including surveillance and monitoring, footage in movies, sports events, inventory verification, and inspection, cargo delivery, communication platforms, rural environment inspection, and disaster response and emergency relief \cite{Xu2018}\cite{Jeong2018}. Such wide-spread applicability is mainly due to aerial vehicles' enormous capabilities in terms of mobility, autonomy, communication, and processing capabilities, in addition to its relatively low cost\cite{Reina2018a}.  

The rapid growth of the Internet of Things (IoT) produces significant challenges in areas such as computer networks, distributed systems, infonomics, and data science. Its effects are already perceived in our society, not only with smartphones that carry multiple sensors but also with devices such as foot pods, heart rate monitors, and connected watches. Those connected devices can collect useful information about an individual and its surroundings.

In many cases, it is difficult or even impossible to connect some devices straight to the Internet or reach wireless sensors and actuator networks (WSAN) directly to their data destination. However, if the smart devices and WSAN can assemble and organize themselves as a Wireless Mesh network (WMNs), and any of the Mesh nodes can serve as an intermediate (i.e., the message relay) for all the other Mesh nodes. The probability of connection and the duration of the connectivity of the whole WMNs with the Internet can be enhanced significantly. 

Typical UAV applications may involve the relaying time-critical data generated from devices on the ground to remote ground stations connected to the Internet \cite{Secinti2018}. UAVs can act as mobile data collectors and delivers, such as connection nodes \cite{Kerrache2018}. Moreover, UAVs can reach remote actuators (E.g., a valve in oil pipelines) to update instruction and temporary control. In such cases, UAVs can provide these connections by visiting the devices periodically and relaying or carrying the data to the proper destination. Recently, there has been growing interest in data collection through groups of collaborative UAVs. \cite{alemayehu2017}\cite{sarmad2017}. 

UAV coordination and their decentralized control are, therefore, timely topics \cite{sarmad2017}. Besides, communication is one of the most significant challenges in designing systems with multiple flying vehicles (a.k.a fleets) and a crucial aspect of cooperation and collaboration \cite{Pigatto2016}. In order to coordinate UAVs in distributed tasks, an inter-UAV communication approach must be effective.

A reliable communication infrastructure among UAVs in collaboration is critical in maintaining this connected network for data relaying tasks \cite{Secinti2018}.  For this reason, reliable communication requires a radio base station close to the site. However, in some cases implementing this infrastructure to provide long-range UAV communication is not possible—for example, in emergency response and relief situations or radio denied environments. Some emergencies, such as earthquakes, could destroy any existing communication infrastructure. In such cases, the incident's location may be uncertain, and obtaining temporary infrastructure equipment (such as a mobile cellphone radio station) may not be immediately possible. These reasons make it impossible to use vehicle-to-infrastructure (V2I) solutions. In these cases, ad-hoc communication plays an important role.

Coordination of UAVs in a distributed manner is a complex task \cite{Garraffa2018} because there is no central node that can have a consistent view of the state of all UAVs (i.e., the instantaneous position, velocity, residual battery level, etc.); furthermore, the ad-hoc communication topology also limits consensus and task sharing. Therefore, such coordination requires effective algorithms to overcome these constraints. Providing a guarantee of effectiveness may threaten the efficiency regarding the amount of collected data or concerning the delay in collecting data from devices by UAVs. UAVs have shown tremendous growth, both in research and applied use \cite{Sayeed2018}. Concerning fleets of UAVs, the current UAV simple model uses a single ground controller to control one or more UAVs \cite{Wu2018a}.
Regarding the control of UAVs, the first issue to resolve is the path plan. For this purpose, the traveling salesman problem (TSP) is widely considered \cite{Huang2019}. Most of these studies focus on using a single UAV in an optimized tour or split the role problem into a smaller set of the same problem to be solved in the same way. It remains challenging to develop cooperative UAVs on area coverage tasks and energy-efficient UAV communication technology \cite{Ruan2018}. Moreover, as we aim to use a fully distributed approach, path planning should be computed and updated in the UAVs, regarding its computational constraints.  

To address the mentioned issues, this research's Section investigates which would be an efficient and reliable approach to coordinate UAVs in a distributed manner in order to connect to IoT devices, sensors, or WSAN through Bluetooth mesh networks (Bmesh). Moreover, we aim to use several fully autonomous UAVs (i.e., not centralized human-controlled) relying only on ad-hoc communication. 

This research aims to validate this hypothesis by extending our current research regarding decentralized UAV fleet control \cite{olivieriIROS17}\cite{olivieriMSWIM17} to handle and map the trade-offs of the use of BMesh with smart devices as the edge of a communication system. 

More specifically, we are developing decentralized and adaptive UAV fleet coordination that relies only on ad-hoc communication that uses only short-range radios from this research. We aim to evaluate how a fleet of autonomous UAVs can interact with ground and nodes in variable locations on the ground to exchange data. 

\subsection{Applications}
\label{sec:applications}

In Incident response, a large set of micro UAVs can be easily carried by a single Minivan and released in the incident vicinity. For example, some incidents, earthquakes, and tsunamis can destroy all communication infrastructure, obliging ad-hoc communication for a while. UAVs can spread a massive amount of sensors on the ground of the affected area and collect data from them.  Furthermore, as many smartphones and intelligent wearable are Bluetooth enabled, after an earthquake, for instance, some UAVs can seek survivors by looking for smartphone radios.

In security and environment monitoring, the visit of points of interest (POI) plays a central role. Critical locations such as hard to reach segments of countries borders or high voltage towers can be easily visited by UAVs to exchange data or pass by a visual inspection. Under environment monitoring, Vast areas usually have some sensors already in place sensor enabling a WSAN or even attached to animals under interest.  In this case, the UAVs visitation in such POI can enable or increase the data exchange. 

In environmental monitoring, vast regions in some types of rainforests have dry seasons, and they are very susceptible to spontaneous or criminal fires that take advantage of the season to destroy the vegetation. In other cases, other tropical forest types can be too dense even for locomotion to a fire outbreak. In either case, it is possible to monitor those vast areas through satellite. However, to stop criminal activities or to combat fire outbreaks, it is necessary to act locally. UAVs can extend fire spot-checking capabilities, assist in prioritizing zones to be treated at the early stages of fire. Moreover, extensive areas can be previously monitored by sensors and visited by UAVs for data collection.

Another possible advantage of using UAVs is taking photographs near the treetops and enabling the identification of illegal lumberjack equipment through the treetops, a fact that would not be possible in high altitude images due to the density of the forest.

Due to its high maneuverable capabilities, UAV can be used in smart cities and forests in diverse contexts. For example, some UAVs can visit each tree of a region of interest to take accurate measures such as detailed images for the tree's life quality or size and high measures. UAVs can also inspect obstructed water runoff areas to prevent flooding.

All the listed possible applications, autonomous UAVs capable of (re) organizing its fleets to divide the overall workload, bring advantages in terms of fail tolerance in terms of UAVs fails or even the better usage of new UAVs entering to the fleet the reinforce the task. Moreover, by UAVs only relying on short-range ad-hoc communication, the entire approach presents more resilience of external interference and independence of external actors. As the UAVs are not supposed to fly close to each other, the short-range radios bring challengers in such coordination.

The problem gets more interesting when we consider unplanned paths for data collection. The UAVs informs the WSAN through a mesh network whenever it updates its route to the network, try to organize itself better to connect with the UAV again in the near future. Whether we use UAVs as a way of connection, there is a difference in whether there is a known coordinate of POIs where we can predict where and/or when there will be a connection with a device and when we cannot. If there is no way to predict which devices will be used for a connection, this leads to some of the algorithms we aim to discuss in this work. These algorithms, distributed and with a certain level of fault tolerance, will be one of our main contributions.

\subsection{GrADyS-A Proposal}
Bluetooth low energy (BLE) devices are capable of being embedded in almost any place. Those devices can exchange messages and collaborate to automate processes, be used inside wearable, stay online connected to sensors left unattended in a forest for long periods, be embed on or bring more comfort to customers that can enjoy plug-and-play home automation. Bluetooth Mesh (Bmesh) allows for simultaneous connections across hundreds of connected devices. 

However, BLE devices have a short-range of communication, and it is not always possible to be used to provide a direct link to the Internet. In section \ref{sec:applications}, we present some application that needs \textemdash or has their usage improved \textemdash some last-mile integration. 

Under these circumstances, we propose an approach to control a fleet of UAVs in a distributed manner capable of interact with WSAN data with BLE devices. Besides, the fleet of UAVs will also rely only on short-range radios without any human supervision and flying away from each other in distances bigger than the radio ranges.

We aim to provide a distributed manner to several small UAVs continuously divide a set of POIs that need recurrent visitation and interaction. We can categorize such points as fire focus points (e.g., hazard borders been photografed) or data exchange points (BLE devices). Such points of interest can also be seen as hot spots suggested by higher overflights. Even more so, in the next phase, mobile sensors may be carried by firefighters attacking the fire or police forces after criminal activity.

\begin{figure}
	\centering
	\includegraphics[width=0.7\columnwidth]{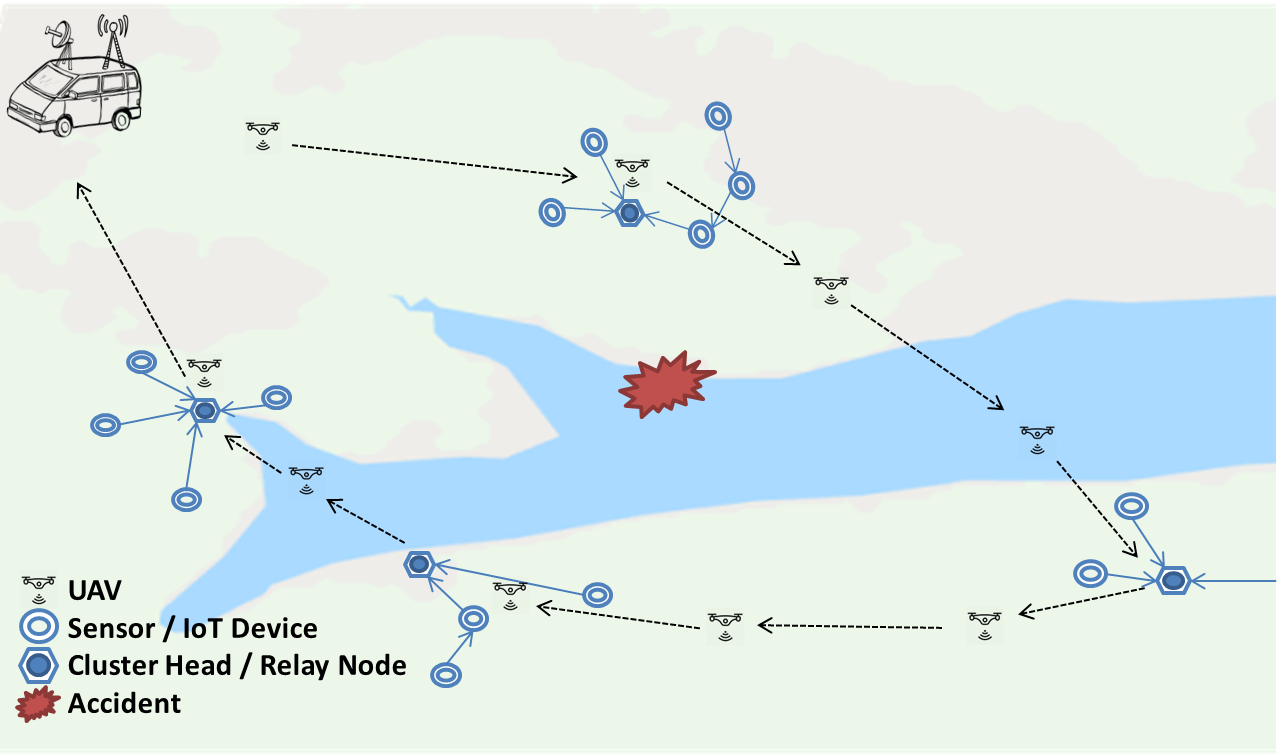}
	\caption{An autonomous fleet of small UAVs launched from a minivan visiting sensors to collect data.}
	\label{fig:dadca_visiting_pois}
\end{figure}

The Figure  \ref{fig:dadca_visiting_pois} show an example of an autonomous fleet of small UAVs launched from a minivan visiting sensors to collect data such as temperature, humidity, and CO gases. The idea is to provide a distributed manner to the UAVs stay continuously retrieve data from the WSAN during a period. During that period, UAV's set will necessarily vary by losing UAVs in operation and receiving reinforcement. The UAV's path is planned by their computational units and updates whenever a UAV enters or leaves the network.

\begin{figure}
	\centering
	\includegraphics[width=0.7\columnwidth]{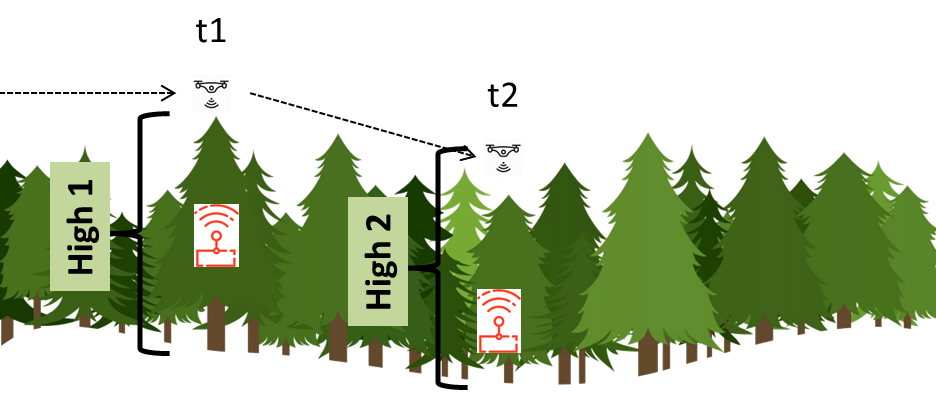}
	\caption{A UAV is varying changing its flight altitude to reach the BLE rage of a device to exchange data.}
	\label{fig:uav_chaging_high}
\end{figure}

The flight altitude of UAVs needs to comply with the ground devices' radio ranges. Due to BLE's range, the UAVs can not always fly over every terrain with the same flight altitude. Our Model will consider a 3D coordinates despite the major related works, as illustrated in Figure \ref{fig:uav_chaging_high}.

\subsection{Planned Real-World Deployment}

In the previous section, regarding the decentralized control of UAVs in the WSAN and WMesh visitation, we intend to evaluate with prototypes of UAV groups, besides the proposed simulations:

\begin{itemize}
	\item How do coordination algorithms behave on real-world flights?
	Are there discrepancies in theoretical communication models and data collection in real WSAN and WMesh networks?
	\item What are the key points that affect UAVs' coordination and data collection?
\end{itemize}

We aim to use BLE, and in most cases, in cities' boundaries. Those radios work under heavy interference due to the extensive usage of applied frequencies. Moreover, the range of such radios is narrow. While simulations bring a fast response in algorithms evaluations, real-world tests can put them under real probation. While the use of BLE may be the subject of radio interference issues, smart cities with its high density of devices operating in the 2.4GHz spectrum, in vast and dense forest areas, its limited range of communication brings the necessity of the flight proximity to collect data.

Tropical rainforests bring unique challenges to the high density of vegetation, its high humidity, and varying water height, all of which make sensor conservation and RF transmission very complicated. The RSSI of the sensor's communication in different forests fluctuates more significantly in the descending trend for each transmits power\cite{dingsensors2016}. Ding et al. observe that the reflection and scattering caused by relatively dense vegetation do not impact the low-power radio link as much as predictable in their studies\cite{dingsensors2016}. In their study, they also found that the complicated forest environments usually lead to high link asymmetry; that is, the qualities of a link on two directions are conspicuously different\cite{dingsensors2016}.

This proposal plans to evaluate this proposal with the Projet IoTrees from the INCT InterSCity\footnote{http://interscity.org} for implementing data collection of sensors disconnected from the Internet with UAVs as a result of the first phase. As a result of the second phase, we aim to implement a flying fleet capable of search and connect to movable devices running BLE services to relay its location and exchange data.

With the joint test with the IoTrees project, it is possible to analyze the same sensor network in more than one application. For example, treetop senors can help locate endangered animals while monitoring environmental conditions that may facilitate forest burning scenarios.

Due to the unique characteristics of some Brazilian forests (e.g., such as density, humidity, and crown height variation), the placement of sensors in the trees themselves is already a problem due to the short range of the radios in focus. We intend to analyze in real experiments related to the IoTrees project:

\begin{itemize}
	\item What would be the optimal point for the sensor positioning in the trees?
	\item How does the positioning of sensors influence the displacement of UAVs as they need to approach treetops and even stay on top of them for a few moments for data exchange?
\end{itemize}

\subsection{Related work}
The Table \ref{table:relatedworks} present some related works. To the best of our knowledge, most related works in this area of research only consider data collection in a single moment, meaning that at some point, a single UAV passes through every WSAN's cluster head or IoT device to collect data once\cite{Nikhitha2018}\cite{Sun2018}. However, certain situations may require data to be collected for a period $T$, such as search-and-rescue missions or wildfire monitoring.

The presented works on Table \ref{table:relatedworks}, in general, do not provide approaches for multi-UAV distributed coordination in data collection. Instead, they either do not use multiple UAVs or use them in isolation. As discussed in \cite{Qadori2017}, further research is required on multi-collector approaches, particularly regarding cooperation.

This proposal explores collaboration among UAVs performing data exchange and aims to compare new approaches and algorithms to the studies mentioned above in multiple UAVs. The collaboration among UAVs involves dynamically resizing each UAV tour upon a UAV malfunction or reinforcement. This resizing process prevents uncovered devices after a UAV leaves the data collection system. It also prevents variable delivery delays. The collaboration is also responsible for forwarding messages between UAVs to reach a ground station or the Internet without the necessity of a UAV displacement to the GS to deliver its collected data. 

\begin{table}[ht]
	\centering
	\caption{Related Works}
	\rowcolors{2}{gray!25}{white}	
	\small
	\resizebox{\textwidth}{!}{\begin{tabular}{llllll} 	
			\toprule
			\rowcolor{gray!50}
			Related Work  & Nodes relation  & \begin{tabular}[c]{@{}l@{}}Work \& Tour \\ approach \end{tabular} & \begin{tabular}[c]{@{}l@{}}Comm. \\approach\end{tabular} & Multi-UAV       & \begin{tabular}[c]{@{}l@{}}UAV \\ Coordination\end{tabular}  \\ 
			\midrule
			
			\textcolor{red}{GrADyS Air} & \textcolor{red}{Data collection }  & 	\begin{tabular}[c]{@{}l@{}}\textcolor{red}{Polynomial and not fixed path}\\\textcolor{red}{planning} \end{tabular} & \textcolor{red}{ad hoc} & \textcolor{red}{Yes, cooperating} & \textcolor{red}{Distributed}  \\
			
			Dios et al.\cite{dios2013} & Data collection  & TSP straight use  & V2I & No & Centralized \\
			
			\begin{tabular}[c]{@{}l@{}}Several researches\\ \cite{Garraffa2018}\cite{Pascarella2013a}\cite{Thakur2013}\cite{Wu2018}\\\cite{Ladosz2017}\cite{Wang2015a} \end{tabular} & Data collection & TSP optimization & V2I & No  & Centralized \\
			
			Jin Wang et al. \cite{Wang2017} & Data collection & \begin{tabular}[c]{@{}l@{}}Choice of CHs by UAV tour\end{tabular} & V2I & Yes, in isolation & Centralized \\
			
			Wang et al. \cite{Wang2015} & Data collection  & FPPWR &  \begin{tabular}[c]{@{}l@{}} V2I \end{tabular}   & No    & Centralized       \\
			
			\begin{tabular}[c]{@{}l@{}} Some researches \\ \cite{Ebrahimi2018},\cite{fu2018} \& \cite{Sekander2018} \end{tabular} & \begin{tabular}[c]{@{}l@{}}  Data collection \end{tabular}  & \begin{tabular}[c]{@{}l@{}} Optimization for area Coverage \\ and relay through 5G \end{tabular} & V2I   & No    & Centralized  \\
			
			Mazayev et al.\cite{Mazayev2016} & \begin{tabular}[c]{@{}l@{}} Data collection with buffer \\and TTL constraints \end{tabular} & \begin{tabular}[c]{@{}l@{}} 
				Tour optimization \\upon constraints \end{tabular} & V2I  & Yes, in isolation  & Centralized \\
			
			Ma et al. \cite{Ma2017} &  \begin{tabular}[c]{@{}l@{}} Opportunistic data collection \\ from mobile sensors\end{tabular}  & Predefined static route & ad hoc & No & Centralized  \\
			
			Berrahal et al. \cite{Berrahal2016}\cite{Berrahal2015}& \begin{tabular}[c]{@{}l@{}}Data collection borders\\ sensoring \& video surveillance \end{tabular}  & Predefined static route & V2I   & Yes, in isolation & Centralized    \\
			
			Jawhar et al.\cite{Jawhar2014} & \begin{tabular}[c]{@{}l@{}} Data collection, but in \\specific sensor distribution (LSN) \end{tabular} & Predefined static route  & Hierarchical V2I & Yes, in isolation & Centralized  \\
			
			Yanmaz et al.\cite{yanmaz2018} & Direct sensoring & mTSP straight use & V2I \& ad hoc & Yes, in isolation & Centralized \\
			
			Zhang et al.\cite{zhang2018b} &  Data dissemination w/ constraints & Optimization for area Coverage & V2I & Yes, but static & Centralized  \\
			
			Sharma et al. \cite{Sharma2016a}& Data dissemination & \begin{tabular}[c]{@{}l@{}}Workload area divided\\ by the number of UAVs\end{tabular}  & V2I \& ad hoc & Yes  & Centralized  \\		
			
			Thammawichai et al.\cite{Thammawichai2018} & \begin{tabular}[c]{@{}l@{}} Pursue a single target or\\ survey an area of interest \\ for data relay \end{tabular}  & Constraints optimization   & Hierarchical V2I  & Yes  & Centralized   \\
			
			\bottomrule
	\end{tabular}}
	\label{table:relatedworks}
\end{table}



\section{GrADyS-G}

The rapid growth of the Internet of Things (IoT) produces significant challenges in areas such as computer networks, distributed systems, infonomics, and data science. Its effects are already perceived in our society, not only with smartphones that carry multiple sensors but also with devices such as footpods, heart rate monitors, and connected watches. Those connected devices can collect useful information about an individual and its surroundings.

As the number of connected devices grows, there will be an increasing need for technologies that can optimize data collection and transmission of sensor data and configuration and control over smart devices. Those technologies need to be efficient regarding battery consumption for the devices that are not connected to a power source. 
By 2021, 25 billion IoT devices are expected to be connected to the Internet, generating a massive data flow volume.

However, when designing software for connected wireless devices, intermittence of connectivity has to be considered, because smart IoT devices may have to be used in places where there is limited or variable wireless radio signal and/or unstable wireless bandwidth. 
This connectivity problem is further complicated if mobility is an intrinsic feature of the IoT system/application, i.e., when dealing with people, vehicles, robots, movable objects, etc.  Thus, in our Research, we are specifically focused on the Internet of Mobile Things (IoMT), where smart IoT devices/objects may not be permanently associated with specific locations, but are movable. In particular, these smart objects, as well as edge devices (gateways/hubs) providing Internet connectivity, may be moved or move autonomously.

In this context, the problem of variable connectivity is significantly increased, as there is a higher uncertainty regarding the place and moment where/when IoT devices will be able to send/receive data through the Internet. 
However, if the smart devices can assemble and organize themselves as a Wireless Mesh network (WMNs), and any of the Mesh nodes can serve as an intermediate (i.e., the message relay) for all the other Mesh nodes. The probability and the duration of connectivity of the whole WMNs with the Internet can be enhanced significantly. This Research aims to validate this hypothesis by extending our current IoMT middleware ContextNet \cite{endler:OJIOT:2018} to handle also Bluetooth Mesh networks of smart devices as the edge of the IoT system. 

Hence, this work will draw upon the ContextNet's scalable cloud-mobile architecture \cite{David2013} and its Mobile Hub component \cite{Talavera:2015,Vascconcelos:DCOSS:2015}, which runs on Android devices and opportunistically connects with nearby Bluetooth Low Energy (BLE) devices, i.e. the WMN nodes.
More specifically, from this Research we are developing a adaptive Bluetooth-based data collection and routing algorithms for WMNs and test them in one or two real-world use cases where mobility is a central characteristic. It may use some concepts of online movement coordination in flying networks [5]. The results obtained with this Research can be applied to different socio-economic sectors such as agriculture, manufacturing, and transportation. 

\subsection{Related Work}
The work \cite{Badawy:WCNC:2009} propose flow control, Routing, and resource allocation algorithms for WMNs considering solar-powered Mesh Nodes. They model the problem as a directed graph of Mesh Nodes and apply algorithms to optimize data flow given battery and routing constraints such as message priority. Their simulation showed that the proposed algorithms might have high computational complexity, suggesting that they would not be suitable for the concept of Mesh IoMT, where we have hundreds of thousands of devices communicating among them.

Data collection can be done by a mobile node connected to the Internet (such as a smartphone or a drone). This node is a roaming internet connection hub. In the ContextNet middleware \cite{endler:OJIOT:2018}, this is called an M-hub (Mobile-Hub)\cite{Talavera:2015,Vascconcelos:DCOSS:2015}. It can connect to nearby sensors (mobile objects) and transmit their data to the Internet. Currently, its approach is to connect to each sensor, one by one, gathering their stored information and relaying it to a gateway that sends it to a processing server.
Another way of collecting data from a network of sensors is, instead of connecting to all nearby mobile objects, the Mobile-Hub connecting to a single local object that would gather data from other mobile objects, using short-range communications (technologies such as Zigbee or Bluetooth). This allows for the mesh mobile node to collect data faster, as it will connect to fewer mobile objects, and allows for the mobile objects to send the data closer to the mobile node (which may increase the overall data transmission speed).

Todtenberg and Kraemer \cite{todtenberg:AHN:2019} published a survey on Bluetooth multi-hop networks. This survey analyzed over 20 years of Research on the topic and involved not only classic Bluetooth technology but also BLE (Bluetooth Low Energy). It showed that over 85\% of the publications from 1999 to today mid-2019 were based on simulations or analytical results or only described Bluetooth multi-hop networks conceptually. Also, several of the publications analyzed by that survey highlight the need for real-world implementations of those types of networks.

Kopják and Sebestyén \cite{Kopjak:SAMI:2018} compared centralized data collection methods in wireless mesh sensor networks without mobility and analyzed their impact on the nodes' battery life. It was tested in food quality and safety scenario using battery-powered nodes connected to temperature sensors.

Adi et al \cite{Adi:SIET:2017} implemented a WMN using Raspberry Pi micro-controllers using Publish/Subscribe to perform data collection in rural areas to collect temperature and humidity data. On the other hand, \cite{DiFrancesco2011} is an extensive survey of WMNs in which mobility is involved. It defined taxonomy for the data collection processes and analyzed data collection works for unmanned aerial vehicles (drones) acting as data collectors.

The work of Djedouboum et al. \cite{Djedouboum:Sensors:2018} studied the current state of data collection in Wireless Mesh Sensor Networks and analyzed its challenges in the context of Big Data. They also discussed data collection challenges when mobility is involved, like contact detection with data collectors, quality of service (QoS), and location detection.

\subsection{Application}
In agriculture, sensors could be deployed to monitor vast amounts of land. Worker's phones or even drones could be used to gather data from those sensors. This could be achieved using a mobile WMN, where sensors can communicate and forward data to the collectors dynamically as they connect to them. Similarly, in manufacturing, a factory's deployed sensors could exchange data to be sent to the web (to a server that controls the factory plant, for example). It is important to notice that they could also send data to other nodes with the nodes' intent to take a quick decision without the information even needing to reach the web.
For example, in transportation, vehicles could communicate with sensors and with other vehicles to exchange information such as navigation data, road density, and planned routes (that can be used to optimize traffic). This could be done using mobile WMNs to quickly propagate information so that other vehicles can take quick action, to change their routes for example.

As in mobile WMNs, the devices will, in most cases, depend on battery power, it is important to create mechanisms that can optimize battery usage. Battery power information could be considered when reorganizing the WMNs routes, and some "battery load balancing" could be designed to equalize power consumption among nodes.
The problem gets more interesting when we consider unplanned paths for data collection. Whether we use drones or phones to collect data, there is a difference in whether there is a known route where we can predict where and/or when there will be a connection to collect data and when we cannot. If there is no way to predict which nodes will be used for data collection, the network will need to quickly reorganize when connections occur. This leads to some of the algorithms we aim to discuss in this work. These algorithms, distributed and with a certain level of fault tolerance, will be one of our main contributions.

\subsection{GrADyS-G Proposal}
Bluetooth Mesh allows for simultaneous connections across hundreds of connected devices. Those devices are able to exchange messages and collaborate to automate processes, increase the efficiency of an industrial plant, or bring more comfort to customers that can enjoy plug-and-play home automation. The goal of Ground Section to design and compare new Mesh routing algorithms for BLE that, together with the ContextNet framework, aim to increase efficiency in data collection and transmission, envisioning a future where IoMT smart devices are everywhere (e.g. in a supermarket, at home, in a gas station, hospital or amusement park, etc.) efficiently transmitting sensor/state data and receiving updates at high speeds. This will be called the ContextNet Adaptive Mesh Extension (ContextNet-AME).
The dynamism of BLE Mesh networking will allow for fast network reconfiguration. In a nutshell, the moment that a hub (smartphone, drone, or another type of data collector) connects to a WMN node, this node will broadcast a request message which will make the other nodes to send them (directly or via other nodes in the Mesh network) the information that they need to be passed on.

We are performing simulations using the OMNET++/INET framework and put them in practice with a very low-cost solution using ESP32 micro-controllers and smartphones to collect data from fuel tank sensors and use smartphones from the gas station workers as hubs that will send data to the ContextNet cloud.

\begin{figure}
	\centering
	\includegraphics[width=\columnwidth]{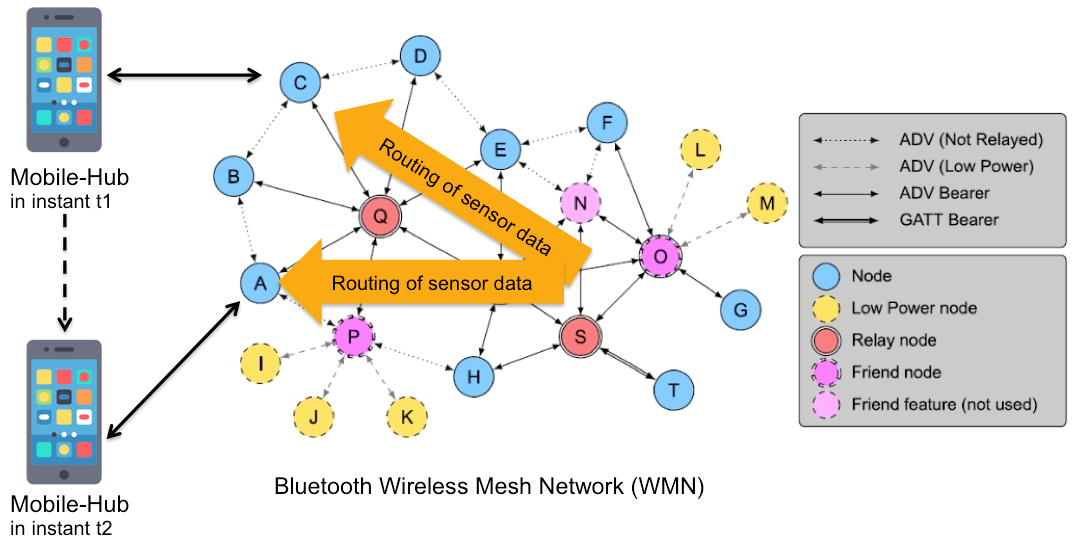}
	\caption{A Bluetooth Wireless Mesh Network routing sensor data to a Mobile Hub that interfaces the WMN at different points.}
	\label{fig:BLE-Mesh}
\end{figure}

\subsection{Definitions}
\begin{description}
	\item[Data collector] (or also referred to as a Mobile Hub) - a smart device with internet access (or access to the final data destination) that is capable of receiving data from other mesh nodes and transmitting it to their destination.
	\item[Low power node (LPN)] is a node (that is usually not connected to the power grid, relying on battery power) that is not always listening for packets and so it relies on FNs to receive them. They request missed packets to FNs when they wake up.
	\item[Friend node (FN)] is a node (that can be powered or not) that is able to receive and acknowledge messages for LPNs during their sleep periods (when they turn their radio off to save power). They transmit received messages on behalf of LPNs upon their request.
\end{description}

According to the Bluetooth Mesh specification2 Friend nodes (FN) of the WMN have a longer – or infinite – lifetime than other nodes. Therefore, these nodes can function as intermediate storage and opportunistic relay nodes for the other "energy-restricted" LPN. These, will typically only awake for communication with some FN during short periods of time so as to save energy.

\subsection{Proposed Routing in the WMNs with Friend Nodes}

For a first version, we are considering WSNs and deployments where there the Mesh has some FNs (friend nodes) which have a constant and infinite power supply and remain static. Thus, the number and position of Friend Nodes will not change, i.e., their set of direct neighbor nodes does not change, and no FN is added or removed. However, the FN's neighbor sets may overlap, so that FNs may distribute the load of re-transmitting sensor data, See Fig. \ref{fig:BLE-Mesh}

\begin{figure}
	\centering
	\includegraphics[width=0.8\columnwidth]{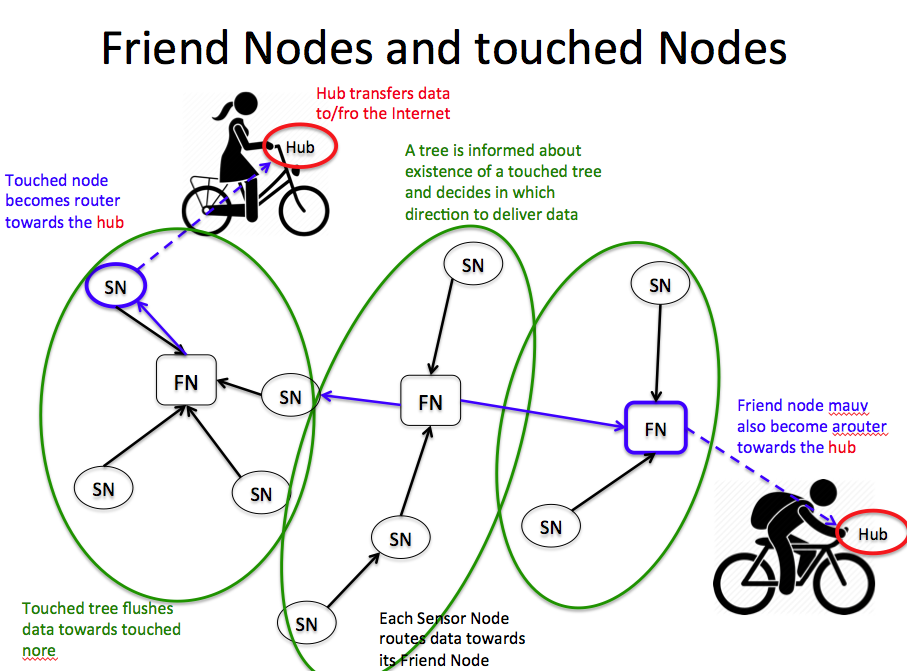}
	\caption{A WMN with 11 nodes and being served simultaneously by two M-Hubs.}
	\label{fig:FN-TN}
\end{figure}

Thus, the basic idea of the routing algorithm for the WMN is as follows:
\begin{itemize}
	\item Once a Data Collector (a.k.a. Mobile Hub) meets and connects with a mesh node, say T1, it informs it of its connectivity status (e.g.: connected to the Internet), asking for sensor data from the Mesh.
	\item At the reception of this message, node T1 initiates a recursive broadcast of a Routing Advertisement towards T1 (RA(T1)), which is propagated from node to node until it reaches all nodes.
	\item Once a node receives this RA(T1), it adds the address of the sending node to its vector mapToReach, indexed by T1.
	\item Since a node may receive a next-hop address towards T1 from several neighbor nodes, it may always update the mapToReach with a new node address if it learns a new route has fewer hops than the previously known route.
	\item Notice that since some mesh nodes behaving as relay nodes may be an ordinary LPN, this broadcast is starting from T1 cannot be instantaneous, but requires the broadcasting node to wait for an awaking period of some of its neighboring nodes, so that the RA(T1) message can be safely received. The time of this waiting is a parameter of the algorithm.
	\item Once RA(T1) is received by a sensor node, including the LPN, it will probe its sensor data, assemble and package with the data, and send it as an acknowledgment to the sending node towards T1.
	\item Once RA(T1) reaches a Friend Node (FN) and it learns how to route messages to the T1, it will: (a) immediately send all accumulated fresh\footnote{Depending on the application, data that is stale for N mins may be considered outdated.} sensor data to T1; and
	(b) set a timeout waiting for its nearby LPNs to wake up and send it to all their current sensor data for delivery. When this timeout expires, it will send the newly received sensor data to T1 over the route it
	has learned through the broadcast.
	\item Notice that all LPNs (one-hop-apart from FNs), which have been requested by an FN to send their sensor data, will also become aware that the sink/destination of this sensor data delivery is T1. The reason is that they should also be able to forward this information mesh network-inwards so that other nodes can register (and optimize) their route towards each contacted node.
\end{itemize}

\subsection{Planned Real-World Deployment}
We plan to use the WMN with this Routing (Sec. 5) in the Project IoTrees from the INCT InterSCity\footnote{http://interscity.org/}, for Smart Cities, for implementing opportunistic "harvesting" of sensor data from Bluetooth ambient sensors (in a WMN -(Wireless Mesh Network)) by biking or pedestrian passers-by.

In Project IoTrees, each tree is instrumented with several sensors (temperature, humidity, SAP flow, etc.) and a microcontroller with Bluetooth LE short-range communication capabilities, that are connected in a WMN. All such nodes periodically collect data from their sensors, store them temporarily in the microcontroller memory, and send the complete data set to neighboring mesh nodes, i.e., nodes within the range of their BLE radio coverage (e.g., 10-50 meters).

\section{GrADyS Interrelationship}

The two project fronts come together in the final objective of the project. At the same time, GrADyS Ground will explore some routing approaches to BLE Mesh to optimize sending data to mobile collectors from sensors on the ground. GrADyS Air will optimize how to collect this data autonomously and cooperatively by UAVs.

This project's main deliverables will be based on the verification and validation experiments of the simulations in the real world. In this way, in addition to the findings related to basic science, they will be compared in tests of applied science. The disparities and cohesion between the experiments' results may bring more factors that support a detailed analysis of tradeoffs.

\section{Results}

GrADyS is a recently started project and already presents preliminary results in both of its research fronts as follows in the following two subsections.

\begin{figure}[t]
	\centering
	\subfigure[Collected data on \textbf{sparse} maps]
	{
		\includegraphics[width=0.31\columnwidth]{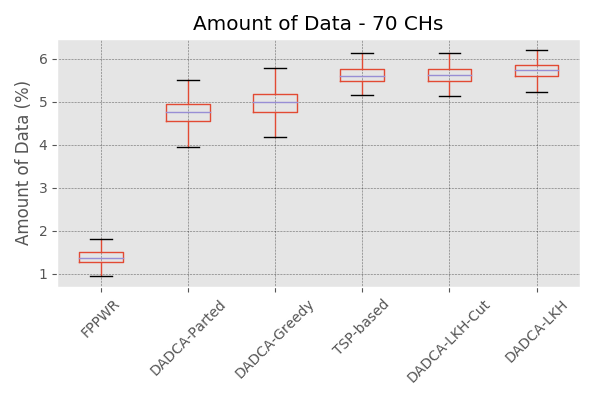}
	}
	\subfigure[Collected data on \textbf{dense} maps]
	{
		\includegraphics[width=0.31\columnwidth]{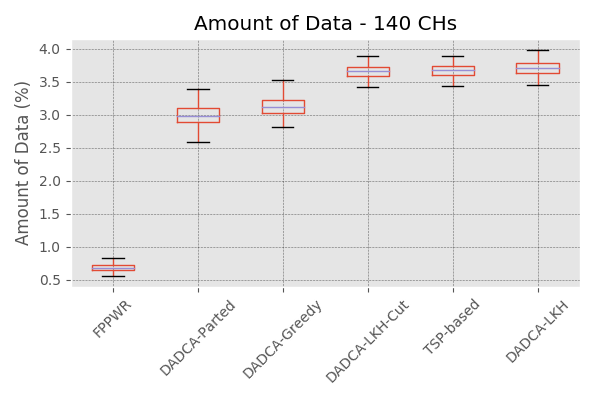}
	}
	\subfigure[Collected data on \textbf{\textit{full}} maps]
	{
		\includegraphics[width=0.31\columnwidth]{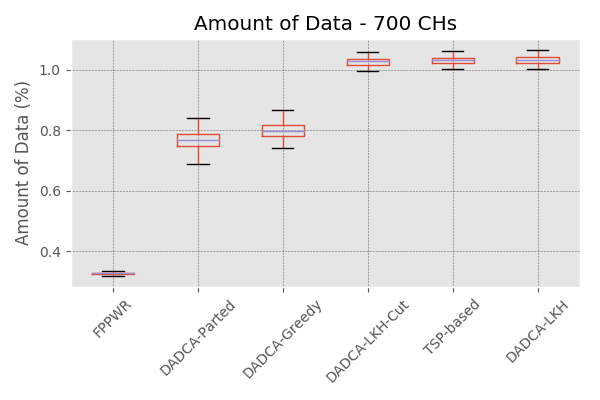}
	}		
	
	\caption{Collected data of all strategies series divided by charts of map densities. The series is crescent ordered by its medians.}
	\label{fig:chpdiscuss:amount:boxplot:all}
\end{figure}

\subsection{GrADyS Air Preliminary Results}

The work developed from the previous \cite{olivieriIROS17} \cite{olivieriMSWIM17} gave rise to the complete work published in \cite{olivieri:JISA:20}. In this work, the variations of the algorithm named DADCA are presented. It was proposed that distributed approaches can be equally or even more efficient than centralized approaches in terms of particular UAV data collection issues. Accordingly, a distributed approach — DADCA — was proposed and evaluated against other strategies (FPPWR\cite{Wang2015} and TSP-based approaches). 

The whole idea that permeates DADCA is that several UAVs receive points of interest and calculate routes that pass through all. Such routes are calculated in the same way in each UAV in such a way that everyone has the same route when receiving the same inputs at eht beginning. However, as UAVs navigate, break, or receive reinforcements, they reorganize themselves by dividing the task of retrieving data from sensors on the ground, generating a stream of message forwarding.  This whole task is done with linear path planning algorithms without trying to calculate the optimal path.

Besides, we evaluated the performance of DADCA variations against conventional strategies with optimized tour approaches. The simulations results reveal that the amount of data collected by DADCA is similar or superior to path optimization approaches by up to 1\%. In our proposed approach, the delay in receiving sensor messages is up to 46\% shorter than in other approaches, and the required processing onboard UAVs can reach less than 75\% of those using optimization-based algorithms.

The Figure \ref{fig:chpdiscuss:amount:boxplot:all} presents some actual results. The vertical axis represents the amount of collected data divided by the generated data (aka a percentage), while the horizontal axis indicates the various strategies names. All charts are ordered based on the series ’medians, from left to right; thus, the farther right a strategy is placed in the charts, the higher are the better are the results.

The results account for a single set of UAVs (eight UAVs) and CH stands for Cluster Heads of a WSN which have data to deliver. All the results, save for the outliers (0.2 \% of the results), are presented in boxplot charts; therefore, 99.8\% of the results are included in the charts. The results appear in Figure \ref{fig:chpdiscuss:amount:boxplot:all}, grouped based on strategy and the density of CHs.

Based on the charts displayed in Figure \ref{fig:chpdiscuss:amount:boxplot:all}, it is possible to cluster the series into three groups. This group formation is more evident in the \textit{full} scenario, as displayed in Figure \ref{fig:chpdiscuss:amount:boxplot:all}. (C). The FPPWR strategy produces isolated results, with the worst positions in all density scenarios across the charts. The DADCA-Naive and DADCA-Parted strategies form the second group achieves significantly better results than FPPWR\cite{Wang2015}. The DADCA-LKH, DADCA-LKH-Cut, and TSP-based strategies aim to create shorter tours and comprise the third group with the best results. The strategy DADCA-LKH achieves the best results because it has shorter tours than the other strategies.

The DADCA-LKH-Cut strategy produces better results than the TSP-based strategy in the sparse scenario, whereas the TSP-based strategy produces the second-best results in the dense and \textit{full} scenarios. These results are due to the efficiency of the DADCA variations in the use of the generated tours. More specifically, the DADCA-LKH-Cut strategy has shorter subtours than the DADCA-LKH strategy and worse results than the TSP-based straight strategy.

\begin{figure}
	\centering
	\includegraphics[width=0.5\columnwidth]{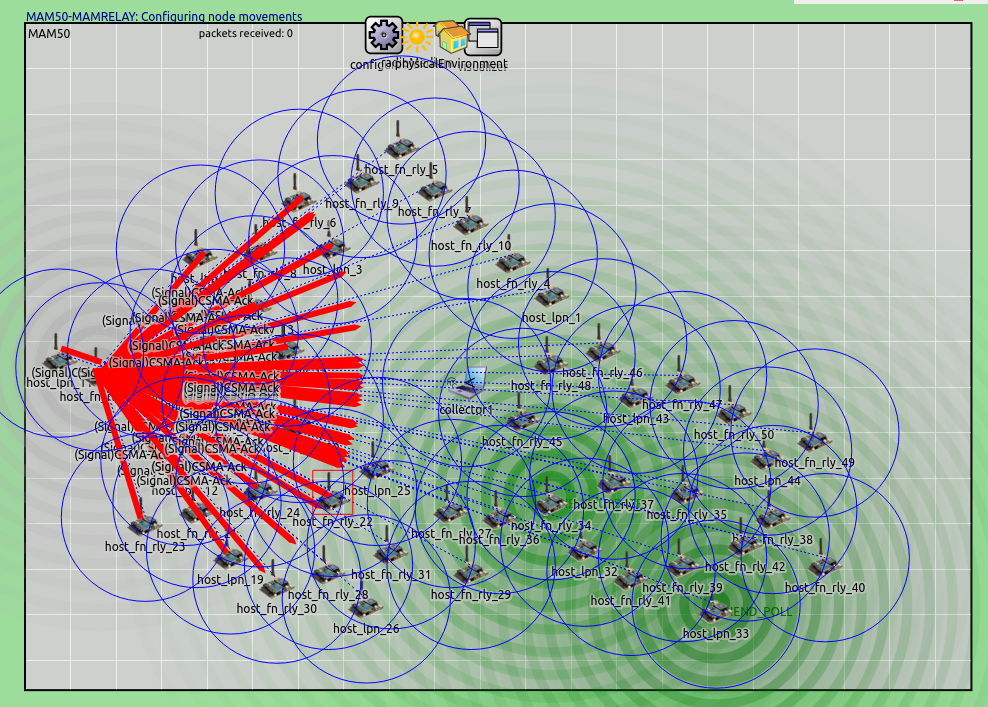}
	\caption{OMNET++/INET simulation visualization}
	\label{fig:mamdiscuss:omnetpp}
\end{figure}

\subsection{GrADyS Ground Preliminary Results}

The ground part of this work consists of using Bluetooth Mesh to form a ground mesh network capable of transmitting sensor data across the ground network until the Mobile-Hub on air. Each Mobile-Hub must transmit a discovery packet to the ground network, every second. The ground network's relay nodes then relay it to all of the network's reachable nodes. Upon receiving such a packet, the nodes will send any sensor data they have to offload. Since Bluetooth Mesh's default Relay algorithm uses a flooding approach without establishing predefined or dynamic/reactive routes, it specifies that such feature can be extended, comparing the original implementation with an extended one designed by us specifically for GrADyS scenarios shown to be the right way to start to evaluate this technology.

Using the OMNET++/INET simulator as shown in figure \ref{fig:mamdiscuss:omnetpp}, we implemented the standard Bluetooth Mesh model, with working Friendship, Low Power, and Relay features. On this application layer, Mobile-Hubs send the discovery packets, and the Mesh network's static nodes send their data upon receiving them. The relay nodes relay the data using the configured algorithm (Bluetooth Mesh or MAM relay) and always relay the discovery packets using a flooding approach. 
The first implementation of the MAM relay node (named $MAM_0$) sends the data only to the last known node with access to a Mobile-Hub (the last to relay a discovery message we received). The initial tests with this approach resulted in less duplicated data being received by the Mobile-Hub, but 63\% less unique data was collected. 

Thus a second version (named $MAM_1$), has the "best route" approach, considering the number of hops of the relayed discovery packets when deciding which node to send the data. This version has a parameter Delta ($\Delta > 0$), which is an expiry time in milliseconds for which we should consider any newer received discovery message as the selected data sink waypoint.

\begin{figure}
	\centering
	\includegraphics[width=0.50\columnwidth]{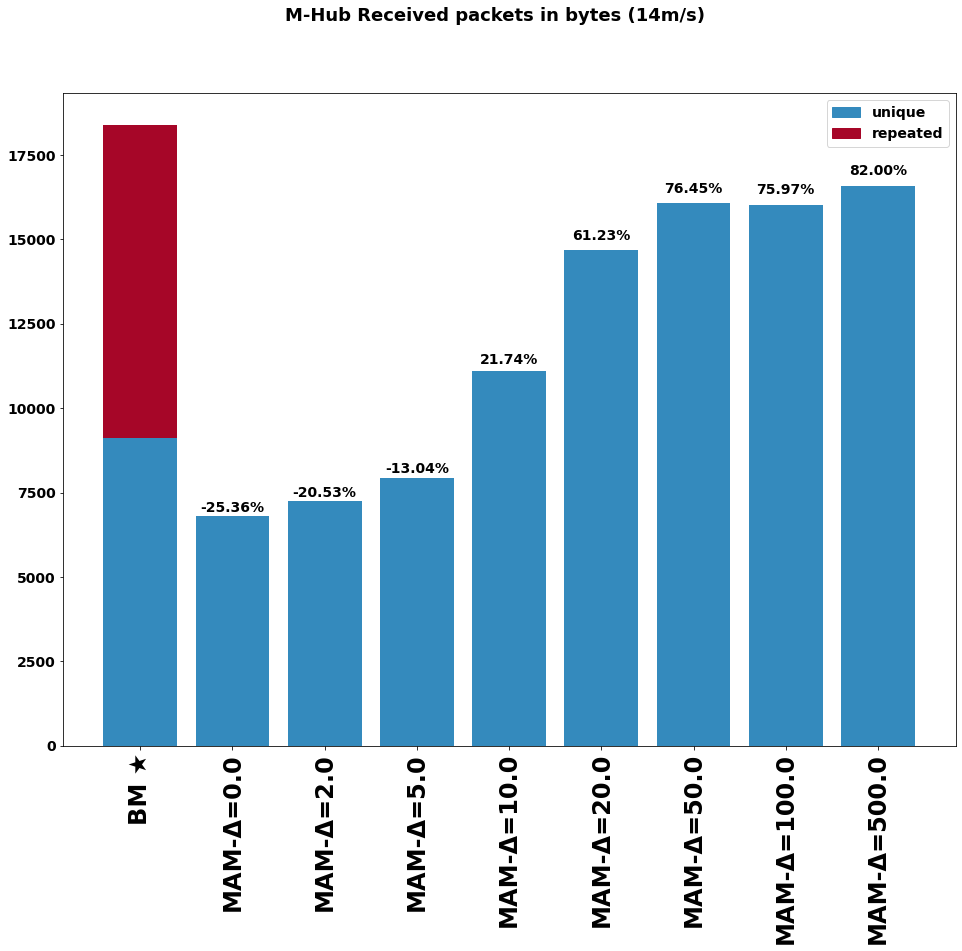}
	\caption{Collected data (in bytes) using Bluetooth Mesh and different MAM parameters using a Mobile-Hub travelling at 14m/s on a circular trajectory.}
	\label{fig:mamdiscuss:data:barplot:all}
\end{figure}

The preliminary results present more than 50\% gain when considering unique messages collected by the Mobile-Hub, after testing with different Delta values, as shown in figure \ref{fig:mamdiscuss:data:barplot:all}.

The simulation used 50 connected ground network nodes and a single Mobile-Hub in a circular trajectory at a constant speed of 14m/s. The network nodes were spread intersecting part of the Mobile-Hub trajectory. The Mobile-Hub took approximately 180s to complete a full pass of its trajectory (that had a radius of 400m).

A detailed paper regarding the MAM routing algorithms is under review in a major conference.

\FloatBarrier
\section{Conclusion}

This research aims to designing, prototyping, and evaluating adaptive distributed algorithms to: 
\begin{itemize}
	\item control a fleets of UAV relaying only on ad hoc communication, with the task of continually collecting sensor data from a wireless mesh network deployed in a isolated region without communication infra-structure, and 
	\item adaptive  routing of (sensor) data in the mesh network according to which mesh node(s) is/are visited by the UAVs at each moment.
\end{itemize}

This paper already presents some relevant initial results related to both GrADyS sub-projects, which will be very helpful when we will design the interaction of the Ground and  Air networks, and when we will deploy either networks in real world scenarios.

We are confident that project GrADyS is opening several very interesting lines of reasearch into dynamic and mobility-aware networks, and with cooperative self-organizaing topologies, as well as on cooperation in systems of heterogeneous networks,  Hence, we believe that the next  steps and future findings will have many exciting results and will point to several use cases in several application fields such as Security/Surveillance, Environmental Monitoring, Smart Cities, Precision Agriculture, and Industry 4.0.

This study was financed in part by AFOSR grant FA9550-20-1-0285.

\bibliographystyle{unsrt}  
\bibliography{bibunificado}  

\end{document}